\documentclass[prb,aps,twocolumn,reprint,amsmath,floatfix,amssymb,showpacs,superscriptaddress]{revtex4-1}
\usepackage{graphicx}
\usepackage{bm}

\makeatletter

\usepackage{babel}
\usepackage{hyperref}
\hypersetup{colorlinks,allcolors=blue}

\newcommand{\ket}[1]{|{#1}\rangle}
\newcommand{\bra}[1]{\langle{#1}|}
\newcommand\numberthis{\addtocounter{equation}{1}\tag{\theequation}}
\newcommand{\aru}{$\alpha$-RuCl$_3$ }
\newcommand{\kggp}{$K\Gamma\Gamma'$ }

\begin{document}
\preprint{APS/123-QED}
\title{Disentangling spin excitation continua in classical and quantum magnets using 2D nonlinear spectroscopy}

\author{Emily Z. Zhang}
\affiliation{Department of Physics, University of Toronto, Toronto, Ontario M5S 1A7, Canada}

\author{Ciar\'an Hickey}
\affiliation{School of Physics, University College Dublin, Belfield, Dublin 4, Ireland}
\affiliation{Centre for Quantum Engineering, Science, and Technology, University College Dublin, Dublin 4, Ireland}

\author{Yong Baek Kim}
\affiliation{Department of Physics, University of Toronto, Toronto, Ontario M5S 1A7, Canada}

\date{\today}

\begin{abstract}

Inelastic neutron scattering (INS) has traditionally been one of the primary methods for investigating quantum magnets, particularly in identifying a continuum of excitations as a hallmark of spin fractionalization in quantum spin liquids (QSLs). However, INS faces severe limitations due to its inability to distinguish between such QSL signatures and similar excitation continua arising from highly frustrated magnetic orders with large unit cells or classical spin liquids. In contrast, two-dimensional coherent spectroscopy (2DCS) has emerged as a powerful tool to probe nonlinear excitation dynamics, offering insights into the underlying mechanisms behind these broad spectral features. In this paper, we utilize classical molecular dynamics (MD) techniques to explore the 2DCS responses of frustrated magnets with dominant Kitaev interactions. Comparing the classical and quantum versions of the pure Kitaev model our results indicate both clear similarities, in the form of sharp line features, and clear distinctions, in the locations of these features and in selection rules. 
Moreover, in the extended \kggp model, we show that the 2DCS response of the Kitaev spin liquid is completely distinct from that of large unit cell magnetic orders, despite both generating a broad continuum in INS. Additionally, we demonstrate the extreme sensitivity of classical 2DCS to thermal fluctuations and discuss the potential significance of quantum coherence in experimental settings. Overall, our work illustrates the potential of 2DCS in resolving the complex physics underlying ambiguous spin excitation continua, thereby enhancing our understanding of the dynamics in these frustrated systems.
\end{abstract}

\maketitle

\section{Introduction}
Detecting quantum spin liquid (QSL) states in experiments has remained a difficult feat given that there are no direct spectroscopic probes for its fractionalized excitations\cite{witczak-krempa_correlated_2014,savary_quantum_2016,zhou_quantum_2017,broholm_quantum_2020}. The dynamical spin correlations captured by inelastic neutron scattering (INS) offer a potential pathway, predicting a continuum of excitations attributed to the fractionalization of the spins in the QSL state\cite{knolle_dynamics_2014, knolle_dynamics_2015}. However, alternative phenomena - such as frustrated magnets with large thermal fluctuations, large unit-cell magnetic orders, or impurity-induced damping of magnons - can similarly manifest broad and featureless INS signals\cite{zhang_spin_2023}. Additionally, a continuum of excitations is not exclusive to quantum states of matter; classical spin liquids, which may arise from materials with large spin-$S$ moments, can produce qualitatively similar excitation continua\cite{samarakoon_comprehensive_2017,samarakoon_classical_2018,zhang_spin_2023}. In principle, the excitation continuum of many quantum spin liquids should exhibit a momentum-dependent lower edge of the continuum, but INS may not have the low energy resolution to observe such a feature. Hence, it is difficult to pinpoint the exact origins of the appearance of a spin excitation continuum with INS, a probe that only accesses the linear response regime. 

Two-dimensional coherent spectroscopy (2DCS) has emerged as a powerful tool to explore the \textit{nonlinear} response in various systems\cite{mukamel_principles_1995,woerner_ultrafast_2013,grishunin_two-dimensional_2023}. Instead of measuring the response to one pulse field as in conventional pump-probe experiments, multiple pump fields are employed, inducing higher-order optical processes, including the generation of higher harmonics as well as the production of sum and difference frequencies. Both theoretical and experimental studies have demonstrated sharp signals in the nonlinear spectroscopic response of various elementary excitations in quantum materials otherwise inaccessible from linear probes\cite{parameswaran_asymptotically_2020,mahmood_observation_2021,nandkishore_spectroscopic_2021,gerken_unique_2022,barbalas_energy_2023,hart_extracting_2023,katsumi_revealing_2023,salvador_principles_2024,wan_resolving_2019,watanabe_exploring_2024,lu_coherent_2017,zhang_terahertz_2024,choi_theory_2020,qiang_probing_2023,brenig_finite_2024}. These include the superconducting Higgs mode in NbN\cite{katsumi_revealing_2023}, Josephson plasmons in layered superconductors\cite{salvador_principles_2024}, spinons in the transverse field Ising chain\cite{wan_resolving_2019,sim_nonlinear_2023,sim_microscopic_2023,watanabe_exploring_2024}, higher harmonic generation from interacting magnons in a canted antiferromagnet\cite{lu_coherent_2017,zhang_terahertz_2024,nelson_upconv_2024}, and Majorana fermions in the Kitaev spin liquid (KSL)\cite{choi_theory_2020,qiang_probing_2023}. Notably, 2DCS on the Kitaev honeycomb model has uncovered sharp diagonal signals separated by 2-flux or 4-flux vison gaps. This naturally leads to the questions of whether 2DCS can illuminate the origins of the spin excitation continua seen with INS, and whether it can be used as a tool to rule out less exotic phenomena behind these measured continua. 

In this work, we provide insights into these questions by studying the 2DCS of classical frustrated magnets with large Kitaev interactions\cite{kitaev_anyons_2006}. It is well known that simulations of the dynamical spin structure factor (DSSF) from the classical Kitaev model produce qualitatively similar results as its quantum counterpart\cite{samarakoon_comprehensive_2017, samarakoon_classical_2018, zhang_spin_2023,bai_magnetic_2019,zhang_dynamical_2019,franke_thermal_2022}. The similarities arise due to the momentum and energy dependence of the spin correlations being largely dictated by the degenerate classical ground state manifold. The primary difference between the classical and quantum KSLs originates from the 2-flux vison gap seen in the quantum model, which may be on the order of 0.6 meV if the Kitaev exchange is about 6 meV as in $\alpha-$RuCl$_3$\cite{kim_crystal_2016}. Such a small gap would be difficult to resolve in INS experiments. Here, we first consider the pure Kitaev model at zero temperature and compare and contrast the 2D nonlinear spectroscopic responses in the classical and quantum models. For the classical model, we compute the 2DCS spectra using molecular dynamics (MD) simulations in conjunction with classical Monte Carlo simulations. Somewhat surprisingly, we observe sharp horizontal, vertical, diagonal, and anti-diagonal line signals in the 2D frequency space. Such sharp line features are characteristic of the quantum KSL, as reported earlier\cite{choi_theory_2020,qiang_probing_2023}, and demonstrate that even classical dynamics can produce sharp ``nonrephasing" and ``rephasing" line features. However, the classical model does not give rise to the two-flux and four-flux gaps responsible for separating and shifting the signals in the quantum case. Furthermore, we see finite responses in various spin polarization channels where signals would be theoretically forbidden in the quantum model. Moving to finite temperature, we observe that the diagonal and anti-diagonal signals are highly sensitive to thermal fluctuations, with a temperature of $T/|K|=0.001$ already enough to wash out these signals.

Lastly, we consider a minimal but realistic \kggp model\cite{rau_generic_2014,sears_magnetic_2015,johnson_monoclinic_2015} for the candidate Kitaev material \aru\cite{plumb__2014,sandilands_scattering_2015,banerjee_proximate_2016,banerjee_neutron_2017,do_majorana_2017,wang_magnetic_2017,banerjee_excitations_2018,zhou_possible_2023}. A recent experiment applying high magnetic fields oriented perpendicular to the honeycomb plane reported a finite region of stability for a possible QSL phase\cite{zhou_possible_2023}. Based on theoretical studies of the \kggp model, this phase corresponds to a quantum paramagnet \cite{gohlke_emergence_2020}, whose status as a spin liquid phase remains to be explored. In the classical limit, this region of the phase diagram corresponds to various competing large unit-cell magnetic orders at zero temperature, which form a thermal ensemble at finite temperature\cite{chern_magnetic_2020}. Remarkably, the DSSF of this ensemble of orders also yields an excitation continuum\cite{zhang_spin_2023}, even though its origins are completely distinct from those from the KSL phase. Despite sharing a qualitatively similar DSSF, we observe strikingly distinct 2DCS responses for the classical KSL versus the ensemble of magnetic orders. The classical KSL shows sharp but smooth lines, in complete contrast to the dense collection of discrete points arising from magnons in the ordered phases. We therefore propose 2DCS as a potentially powerful probe for the underlying excitations of the putative QSL phase in \aru under an out-of-plane magnetic field. 

The rest of this paper is organized as follows. Sec.~\ref{sec:methods} outlines the 2DCS formalism, as well as the numerical methods used to compute the 2DCS. Sec.~\ref{sec:kitaev} presents our results for the pure Kitaev model, with a review of the results of the quantum model in \ref{sec:kitaev}.A, along with the zero and finite temperature results in \ref{sec:kitaev}.B and C. We present the 2DCS results for various ordered phases with spin excitation continua in Sec.~\ref{sec:2dspec} and compare it to the pure Kitaev model. Finally, Sec.~\ref{sec:discussion} summarizes our key findings and points out some open questions for future studies.

\section{Model and Methods}\label{sec:methods}

\begin{figure}[t!]
    \centering
    \includegraphics[width=0.9\columnwidth]{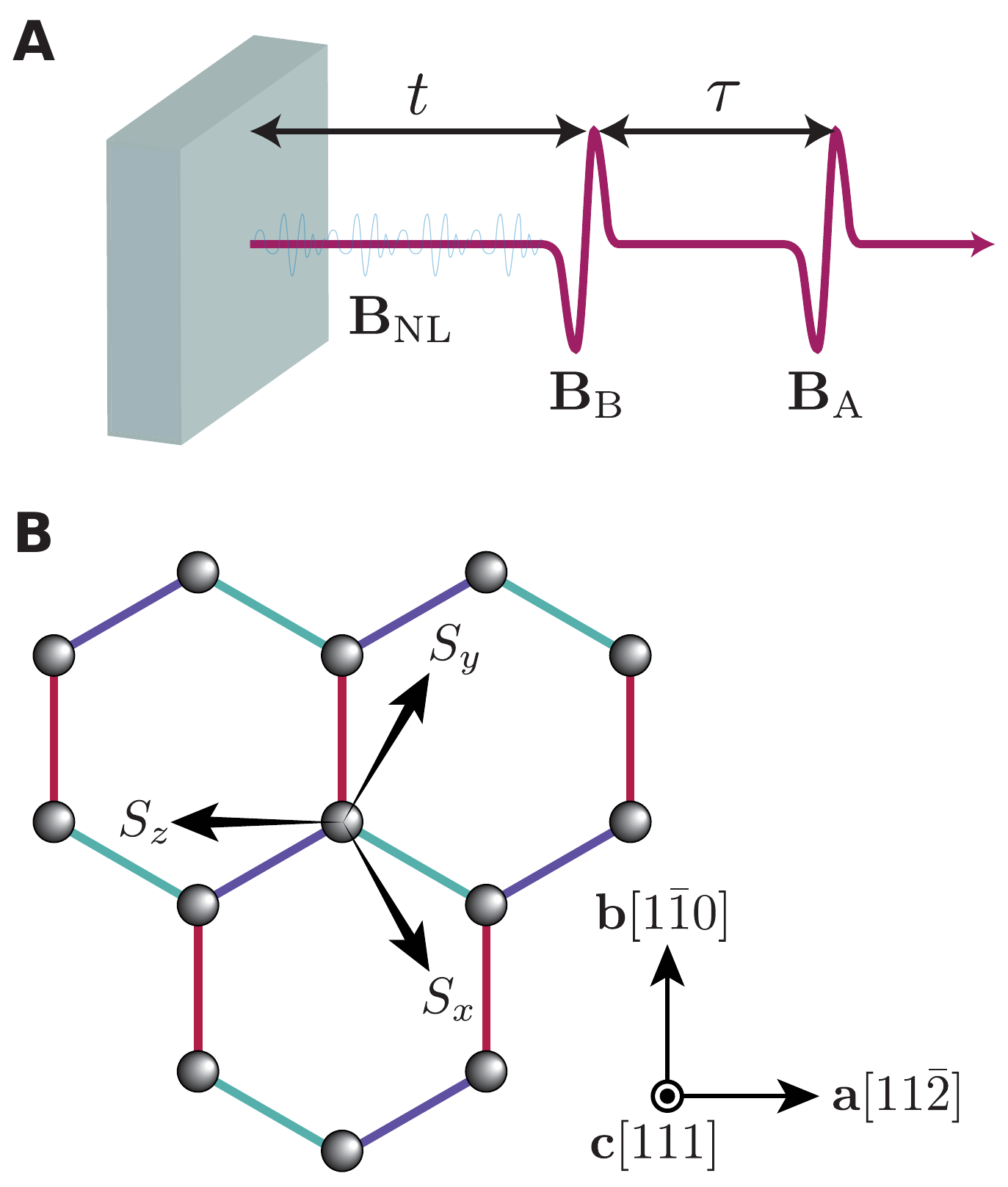}
    \caption{\textbf{A,} Schematic of the 2DCS experiment with incident pulses $\textbf{B}_A$ and $\textbf{B}_B$ separated by time $\tau$. The signal field $\textbf{B}_{NL}=\textbf{B}_{AB}-\textbf{B}_{A}-\textbf{B}_{B}$ is measured at a time $t$ after the second pulse. The incident pulses are polarized in the local $z$ direction, depicted in \textbf{B}.  The spin axis $(S_x, S_y, S_z)$ is coming out of the plane and the global axis $(a,b,c)$ is shown in the basis of the local coordinates. }
    \label{fig:schematic}
\end{figure}

Throughout this paper, we consider the general \kggp model with a Zeeman coupling given by 
\begin{align}\label{eq:ham}
    H=\sum_{\langle ij\rangle\in\lambda}\mathbf{S}_i^TH_\lambda\mathbf{S}_j
-\mathbf{h}^T\sum_i\mathbf{S}_i
\end{align}
where 
\begin{align}
    H_{x}=\left[\begin{matrix}K & \Gamma' & \Gamma'\\
                \Gamma' & 0 & \Gamma\\
                \Gamma' & \Gamma & 0
                \end{matrix}\right],
    H_{y}=\left[\begin{matrix}0 & \Gamma' & \Gamma\\
                \Gamma' & K & \Gamma'\\
                \Gamma & \Gamma' & 0
                \end{matrix}\right],
    H_{z}=\left[\begin{matrix}0 & \Gamma & \Gamma'\\
                \Gamma & 0 & \Gamma'\\
                \Gamma' & \Gamma' & K
    \end{matrix}\right],
\end{align}
and $\textbf{h}$ is an external magnetic field. To study the nonlinear response of this Hamiltonian, we consider two linearly polarized and spatially uniform incident pulses $\mathbf{B}_A$ and  $\mathbf{B}_B$ separated by a delay time $\tau$, with the magnetization measured at a measurement time $t$ after the second pulse, see Fig.~\ref{fig:schematic}A. The pulses linearly couple to the local moments and modify the Hamiltonian by

\begin{align}\label{eq:totham}
    H_{\text{tot}}(t^\prime) = H - (\mathbf{B}_A(t^\prime)+\mathbf{B}_B(t^\prime))\cdot\sum_i\mathbf{S}_i,
\end{align}
inducing a transient time-dependent magnetization $\mathbf{M}_{AB}(t,\tau)$. The nonlinear magnetization can then be extracted by subtracting off the leading contributions from the linear response,
\begin{align}\label{eq:mnl}
    \mathbf{M}_{NL}(t,\tau)=\mathbf{M}_{AB}(t,\tau)-\mathbf{M}_{A}(t,\tau)-\mathbf{M}_{B}(t,\tau),
\end{align}
where $\mathbf{M}_{A}$ and $\mathbf{M}_{B}$ are the magnetizations of the system when only pulse A and B are present, respectively. Finally, the 2D nonlinear spectroscopic responses can be obtained by Fourier transforming in $t$ and $\tau$, resulting in $\mathbf{M}_{NL}(\omega_t,\omega_\tau)$. For the rest of this paper, we take $\mathbf{B}_A$, $\mathbf{B}_B$ to be polarized in the local $z$ direction, while $\mathbf{h}$ is applied in the global $c$ direction, or $[111]$ in the local frame (see Fig.~\ref{fig:schematic}B). 

To simulate the dynamics induced by the incident pulse fields, we first study the static Hamiltonian in Eq.~\ref{eq:ham} using classical Monte Carlo methods, in which we treat the spins as vectors $\mathbf{S}_i=(S_i^x,S_i^y,S_i^z)$. We then time evolve the resulting spin configurations in the presence of the pulse fields using the Landau-Lifshitz-Gilbert (LLG) equations of motion\cite{lakshmanan_fascinating_2011-1}, 
\begin{align}\label{eq:llg}
    \frac{d}{d t^\prime} \mathbf{S}_{i}=-\mathbf{S}_{i} \times \frac{\partial H_{\text{tot}}(t^\prime)}{\partial \mathbf{S}_{i}},
\end{align}
where $H_{\text{tot}}(t^\prime)$ is given by Eq.~\ref{eq:totham}. Using this method, we can directly compute Eq.~\ref{eq:mnl}, and the 2DCS is obtained by performing a 2D discrete Fourier transform. More details of the numerical techniques can be found in Appendix \ref{sec:appmethods}. 

\section{Kitaev model}\label{sec:kitaev}
First, we focus on the pure Kitaev limit, where we can directly compare our simulations on the classical model to the results of the exactly solvable quantum model. For the purposes of comparison, we provide a brief overview of the expected 2DCS signals for the Kitaev spin liquid before we present our MD results. 

\subsection{Quantum Model}

\begin{figure*}[t!]
    \centering
    \includegraphics[width=1.0\textwidth]{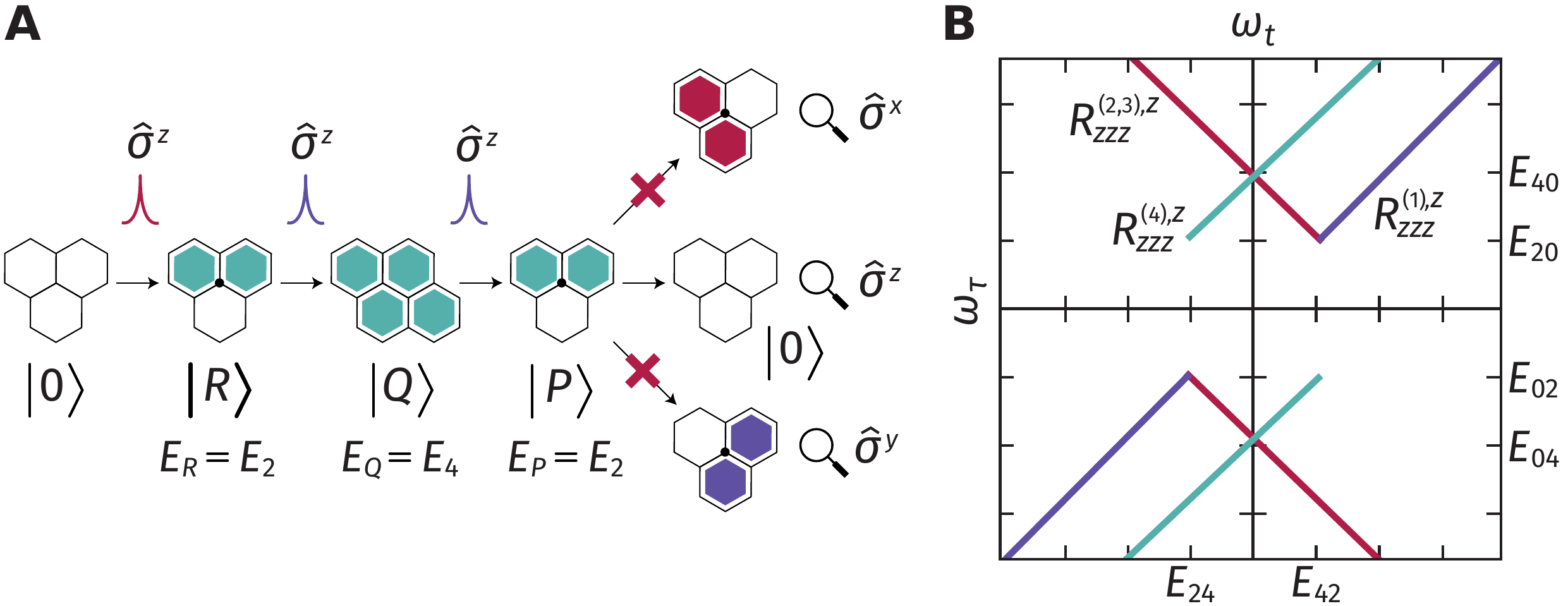}
    \caption{\textbf{A,} Illustrative example of the flux selection rules for the $\chi^{(3),z}_{zzz}(t,0,\tau)$ nonlinear susceptibility. Each $\hat{\sigma}^z$ operator creates or destroys a flux pair across the $z$-bond. The red pulse corresponds to the incident A pulse, while the two purple pulses correspond to B pulses that interact twice. \textbf{B,} Sketch of the two-dimensional Fourier spectrum of the third-order susceptibility $\chi^{(3),z}_{zzz}(t,0,\tau)$. The contributions of each four-point correlation functions obtained by the flux selection rules are indicated. Note that $R^{(3),z}_{zzz}(t,0,\tau)=R^{(2),z}_{zzz}(t,0,\tau)=R^{(2,3),z}_{zzz}$. On the axis ticks, $E_{ab}=E_a - E_b$. Only the flux energies and not the matter fermion energies are shown here for brevity. }
    \label{fig:R3}
\end{figure*}
The induced nonlinear magnetization, $\mathbf{M}_{NL}$, for the quantum model can be obtained by calculating
\begin{align}
\notag    M_{NL}^z(t,\tau)/2N&=\chi^{(2),z}_{zz}(t,\tau)B_A^zB_B^z \\
      \notag              &+\chi^{(3),z}_{zzz}(t,\tau,0)B_A^zB_A^zB_B^z\\
                    &+\chi^{(3),z}_{zzz}(t,0,\tau)B_A^zB_B^zB_B^z+\mathcal{O}(B^4)
\end{align}
where the nonlinear susceptibilities can be computed using time-dependent perturbation theory \cite{mukamel_principles_1995}. For the $B_A^zB_B^z$ pulse configuration, the second order susceptibility $\chi^{(2),z}_{zz}(t,\tau)$ is zero\cite{choi_theory_2020}, and thus the third order susceptibilities fully dictate the behavior of $\mathbf{M}_{NL}$. $\chi^{(3),z}_{zzz}(t_3,t_2,t_1)$ is computed from the four-point correlation functions $R_{zzz}^{(l),z}$, given by 
\begin{align}
    \chi^{(3),z}_{zzz}(t_3,t_2,t_1) = \frac{1}{N}\textrm{Im}\left[ \sum_{l=1}^4 R_{zzz}^{(l),z} (t_3,t_2,t_1) \right]
\end{align}
where 
\begin{align*}
    R_{zzz}^{(1),z}(t_3,t_2,t_1) &
    = \langle \hat{M}^z(t_3+t_2+t_1)\hat{M}^z(t_2+t_1)\hat{M}^z(t_1)\hat{M}^z\rangle \\
    R_{zzz}^{(2),z}(t_3,t_2,t_1) &
    = \langle \hat{M}^z\hat{M}^z(t_2+t_1)\hat{M}^z(t_3+t_2+t_1)\hat{M}^z(t_1)\rangle \\
    R_{zzz}^{(3),z}(t_3,t_2,t_1) &
    = \langle \hat{M}^z\hat{M}^z(t_1)\hat{M}^z(t_3+t_2+t_1)\hat{M}^z(t_2+t_1)\rangle \\
    R_{zzz}^{(4),z}(t_3,t_2,t_1) &
    = \langle \hat{M}^z(t_1)\hat{M}^z(t_2+t_1)\hat{M}^z(t_3+t_2+t_1)\hat{M}^z\rangle.
\end{align*}

Using the resolution of identity, we can rewrite these four-point correlation functions in Lehmann representation. For example, $R_{zzz}^{(3),z}(t,0,\tau)$ can be written as 

\begin{align*}\label{eq:r3}
    R_{zzz}^{(3),z}&(t,0,\tau) =\sum_{jklm}\sum_{PQR} e^{-i(E_R-E_Q)t}e^{+i(E_P-E_0)\tau} \\
     & \times \bra{0}\hat{\sigma}_m^z\ket{P}\bra{P}\hat{\sigma}_l^z\ket{Q}\bra{Q}\hat{\sigma}_k^z\ket{R}\bra{R} \hat{\sigma}_j^z\ket{0} \numberthis 
\end{align*}
where $j,k,l,m$ are site indices and $P,Q,R$ label the energy eigenstates. Here, $\ket{0}$ is the ground state representing the zero flux and zero matter sector. The action of the local spin operators $\hat{\sigma}_j^x$,$\hat{\sigma}_j^y$,$\hat{\sigma}_j^z$ on $\ket{0}$ represents the creation of two fluxes across the $x$, $y$, or $z$ bonds respectively connected to site $j$, while also creating a local excitation in the matter sector. An illustration of a possible $\chi_{zzz}^{(3),z}(t,0,\tau)$ pathway is shown in Fig.~\ref{fig:R3}A, and is an example of how one could deduce the energies of the intermediate excited eigenstates in the flux sector. For instance, the $A$ pulse polarized in the $z$ direction excites a matter fermion and two fluxes across the $z$ bond. Then, the $B$ pulse interacts twice, first creating a four-flux configuration, then annihilating a flux pair and returning to the two-flux sector. A measurement in the $z$ direction finally returns the two-flux state to the vacuum, obeying the cyclic property of the trace in Eq. \ref{eq:r3}. Notably, trying to measure in the $\hat{\sigma}^x$ and $\hat{\sigma}^y$ channels for the $\mathbf{B}_A \parallel \hat{z}$, $\mathbf{B}_B \parallel \hat{z}$ pulse configuration will not yield a finite signal as there lacks an operation that would return us to the flux-free ground state. This principle gives rise to the selection rules in the quantum model: given two incident pulses in the $z$ direction, the only finite component of the magnetization is in the $z$ direction. Furthermore, the same principles dictate that the second order susceptibility must be zero with this pulse configuration.

Equipped with the energies of the intermediate eigenstates, we can roughly predict where the signals should appear in 
$\chi^{(3),z}_{zzz}(\omega_t,0,\omega_\tau)$, a sketch of which is shown in Fig.~\ref{fig:R3}B. Note that this sketch is an illustration of the constraints on $\chi^{(3),z}_{zzz}(\omega_t,0,\omega_\tau)$ rather than an actual calculation of the matrix elements. The main features from each correlator are the diagonal and antidiagonal signals separated by the two-flux or four-flux gaps. Furthermore, the $R_{zzz}^{(2,3),z}$ and  $R_{zzz}^{(4),z}$ signals are offset in $\omega_t$ and $\omega_\tau$. Additionally, as reported earlier\cite{choi_theory_2020}, the $\chi^{(3),z}_{zzz}(\omega_t,\omega_\tau,0)$ channel (obtained from Fourier transforming $\chi^{(3),z}_{zzz}(t,\tau,0)$) gives rise to horizontal and vertical signals with offsets in $\omega_t$ and $\omega_\tau$ by two-flux or four-flux gaps (not shown here). With these features, along with the selection rules described above in mind, we will compare the results from the quantum model \cite{choi_theory_2020,qiang_probing_2023} to our MD results in the following subsections. 

\begin{figure*}[t!]
    \centering
    \includegraphics[width=1.0\textwidth]{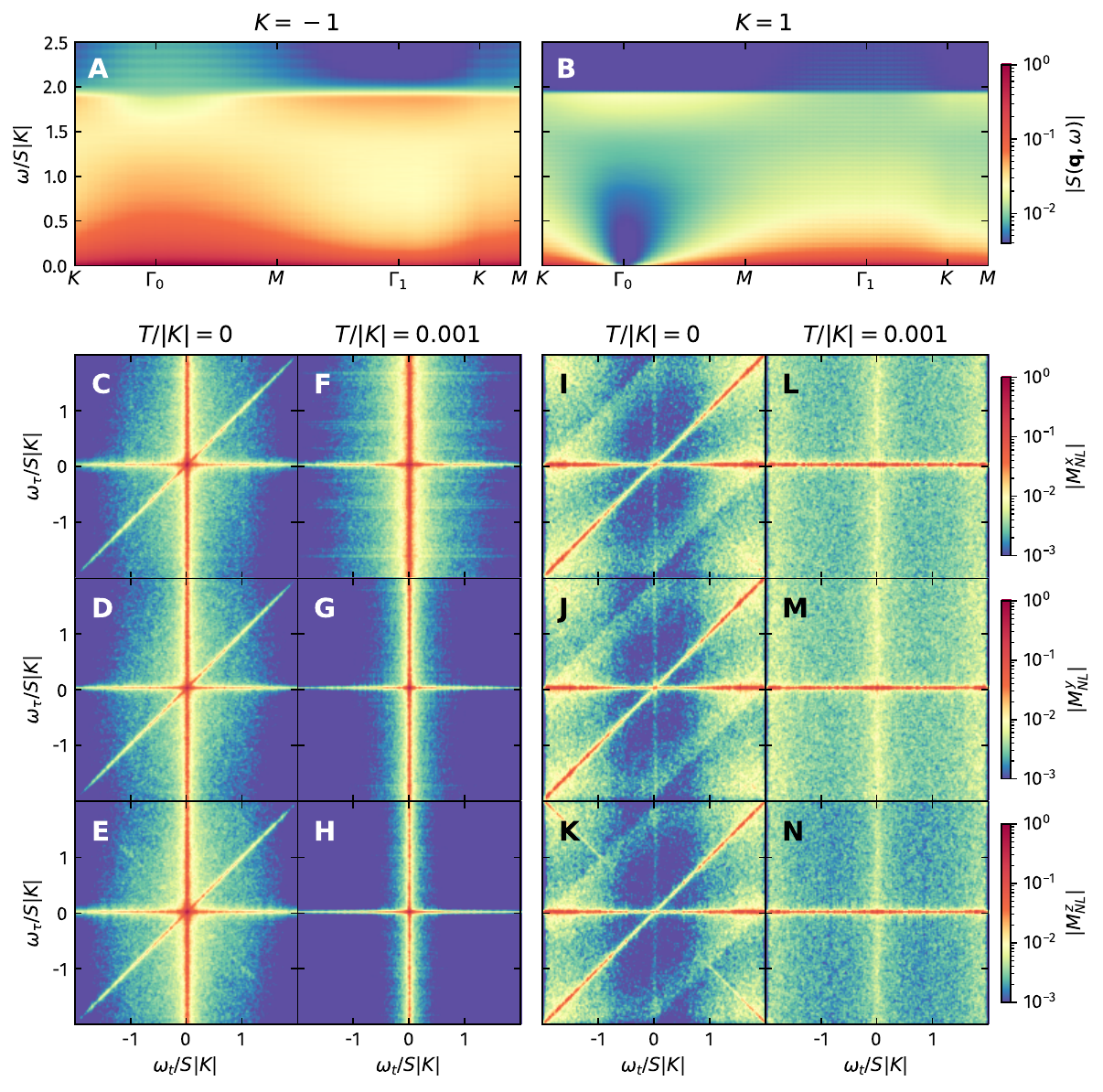}
    \caption{Dynamical spin structure factor (\textbf{A,B}) and two-dimensional coherent spectroscopy (\textbf{C}-\textbf{N}) for the pure Kitaev models with $K=-1$ and $K=1$.  \textbf{A} and \textbf{B} are replotted from \citet{zhang_spin_2023}, and were performed at $T/|K|=0.001$. $|M_{NL}^{x}(\omega_t, \omega_\tau)|$ (\textbf{C}-\textbf{L}), $|M_{NL}^{y}(\omega_t, \omega_\tau)|$ (\textbf{D}-\textbf{M}), and $|M_{NL}^{z}(\omega_t, \omega_\tau)|$ (\textbf{E}-\textbf{N}) are shown for $T/|K|=0$ and $T/|K|=0.001$. All plots were normalized to their respective maximum intensities. Before taking the Fourier transform, a Gaussian filter of $e^{-\eta(t^2+\tau^2)}$ was applied, with $\eta=10^{-6}$. }
    \label{fig:purekitaev}
\end{figure*}

\subsection{Classical Model: Zero Temperature}
We first simulate the classical Kitaev model at zero temperature in order to directly compare against the quantum model. For ease of reference, we plot the dynamical spin structure factors for the ferromagnetic and antiferromagnetic Kitaev model from \citet{zhang_spin_2023} in Fig.~\ref{fig:purekitaev}A and B, which qualitatively reproduce the results of the quantum model except for the two-flux gap\cite{samarakoon_comprehensive_2017, yoshitake_majorana-magnon_2020}. The 2DCS for $K=-1$ and $K=1$ for three measurement channels ($x$, $y$, and $z$) are shown in Fig.~\ref{fig:purekitaev}C-E and Fig.~\ref{fig:purekitaev}I-K, respectively. In both cases, we see sharp diagonal nonrephasing signals in each measurement channel, and faint but sharp antidiagonal rephasing signals in the $z$ measurement channel (in Figs.~\ref{fig:purekitaev}E and K). The sharp diagonal signal is reminiscent of the $R^{(1),z}_{zzz}$ contribution from Fig.~\ref{fig:R3}B in the quantum model, apart from the 2-flux gap. The horizontal and vertical signals are comparable to the contributions from the $\chi^{(3),z}_{zzz}(\omega_t,\omega_\tau,0)$ susceptibility, where the $A$ pulse interacts twice\cite{choi_theory_2020}. 

\begin{figure}
    \centering
    \includegraphics[width=1.0\columnwidth]{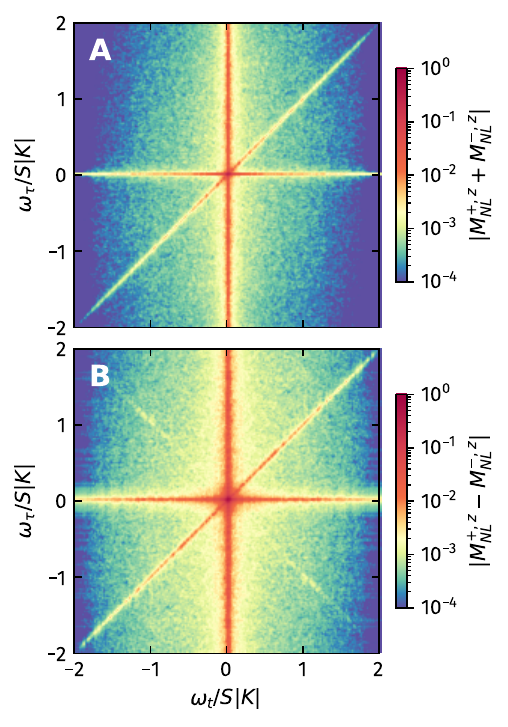}
    \caption{\textbf{A,} Even and \textbf{B,} odd decomposition of the 2DCS for the pure $K=-1$ model at zero temperature. $\textbf{M}^{+}_{NL}$ ($\textbf{M}^{-}_{NL}$) refers to the nonlinear response with two pulses polarized in the $+\hat{\textbf{z}}$ ($-\hat{\textbf{z}}$) direction. Before taking the Fourier transform, a Gaussian filter of $e^{-\eta(t^2+\tau^2)}$ was applied, with $\eta=10^{-6}$. }
    \label{fig:evenodd}
\end{figure}

 We can immediately notice a number of discriminating features between the classical and quantum limits of the model. Firstly, since there is a finite signal in each measurement channel, the classical model does not obey the flux selection rules present in the quantum model (similar in spirit to the linear response case, in which off-diagonal $\chi^{(1),\alpha}_\beta$ vanish in the quantum model but are finite in the classical model). Secondly, although the classical model is able to produce sharp horizontal, vertical, diagonal, and antidiagonal signals, there is no offset of the signals by two- or four-flux gaps, as seen in Fig.~\ref{fig:R3}B. In other words, the classical model cannot distinguish between the diagonal $R^{(1),z}_{zzz}$ and shifted diagonal $R^{(4),z}_{zzz}$ contributions. Note though that both the flux selection rules and hard flux gap discussed here are unique to the exactly solvable Kitaev model. Upon including perturbing interactions, these properties no longer necessarily apply, resulting in more similar quantum and classical 2DCS responses. 

One can further decompose the nonlinear response into its even and odd components, $\textbf{M}_{NL}=\textbf{M}_{NL,\text{even}}+\textbf{M}_{NL,\text{odd}}$, where $\textbf{M}_{NL,\text{even(odd)}}$ consists of the even (odd) order responses. By performing the simulation with the two pulses co-polarized in either the $\pm\hat{\textbf{z}}$ directions, the even/odd-order responses can be separated into
\begin{align}
    \textbf{M}_{NL,\text{even}}&=\frac{1}{2}(\textbf{M}_{NL}^{+}+\textbf{M}_{NL}^{-})\\
    \textbf{M}_{NL,\text{odd}} &=\frac{1}{2}(\textbf{M}_{NL}^{+}-\textbf{M}_{NL}^{-}),
\end{align}
where $\textbf{M}_{NL}^{\pm}$ are the nonlinear magnetizations with both pulses polarized in the $\pm\hat{\textbf{z}}$ direction. We present the separation of the even- and odd-order responses for the pure Kitaev model at zero temperature in Fig.~\ref{fig:evenodd}. As mentioned earlier, the $\chi^{(2),z}_{zz}(t,\tau)$ (and higher even-order) response should be exactly zero for the quantum model due to the flux selection rules\cite{choi_theory_2020,qiang_probing_2023}. However, Fig.~\ref{fig:evenodd}A shows a finite signal in the even-order response, again consistent with the fact that the classical model does not obey these selection rules. 
On the other hand, we see an antidiagonal signal appear only in the odd response shown in Fig.~\ref{fig:evenodd}B. At leading order, this signal is unique to the third order $R^{(2,3),z}_{zzz}$ ``rephasing" response in the quantum model, meaning that even at the classical level, we observe the sharp line features at the correct order. 

Overall, the classical model is able to capture global features of the 2DCS response of the quantum spin-$1/2$ model, including sharp line signals and the correct leading order response. In order to properly differentiate between the two limits, one must utilize the unique nonlinear 2DCS signatures arising from the flux excitations (or visons) of the quantum model.

\subsection{Classical Model: Finite Temperature}
Next, we present the 2DCS for the pure Kitaev model at $T/|K|=0.001$, shown in Figs.~\ref{fig:purekitaev}F-H for $K=-1$ and Figs.~\ref{fig:purekitaev}L-N for $K=1$. Although the horizontal and vertical signals survive, the diagonal nonrephasing and antidiagonal rephasing signals are completely absent for both models, implying that these signals are extremely sensitive to thermal noise. The reasoning for the sensitivity to temperature in the classical model is as follows. At zero temperature, the spin configurations are in the lowest energy configuration (forming a degenerate classical ground state manifold), meaning that the spins are almost perfectly aligned towards the local effective field $\textbf{H}_\text{eff}=\frac{\partial H}{\partial \textbf{S}_i}$, which appears in Eq.~\ref{eq:llg}. When performing the time evolution using the LLG equations, the right hand side of Eq.~\ref{eq:llg} is only finite when there are time-dependent pulse fields present, thus the only spin dynamics are due to the presence of the pulses. 
In contrast, as the temperature is increased, the spins become misaligned with the local field, meaning spin dynamics are already present due to thermal fluctuations even without the pump fields. While these thermally-induced fluctuations are essential for the emergent spin excitation continuum in the DSSF, they may wash out the coherent nonlinear dynamics induced by the pulsed fields in 2DCS. 
A more detailed discussion on the decoherence of the spectroscopy is found in Appendix \ref{sec:appthermal}. On the other hand, the diagonal and antidiagonal signals may be more robust in the quantum model since the quantum coherent signals may be less sensitive to thermal fluctuations. For the simpler case of the 1D transverse-field Ising model, it was shown that such signals can survive at finite temperatures\cite{watanabe_exploring_2024}. Hence, the investigation of the effects of temperature on the 2DCS signals in the quantum Kitaev model is a fruitful topic for future studies. 

\section{Spin excitation continua and 2D spectroscopy 
}\label{sec:2dspec}

\begin{figure*}[t!]
    \centering
    \includegraphics[width=1.0\textwidth]{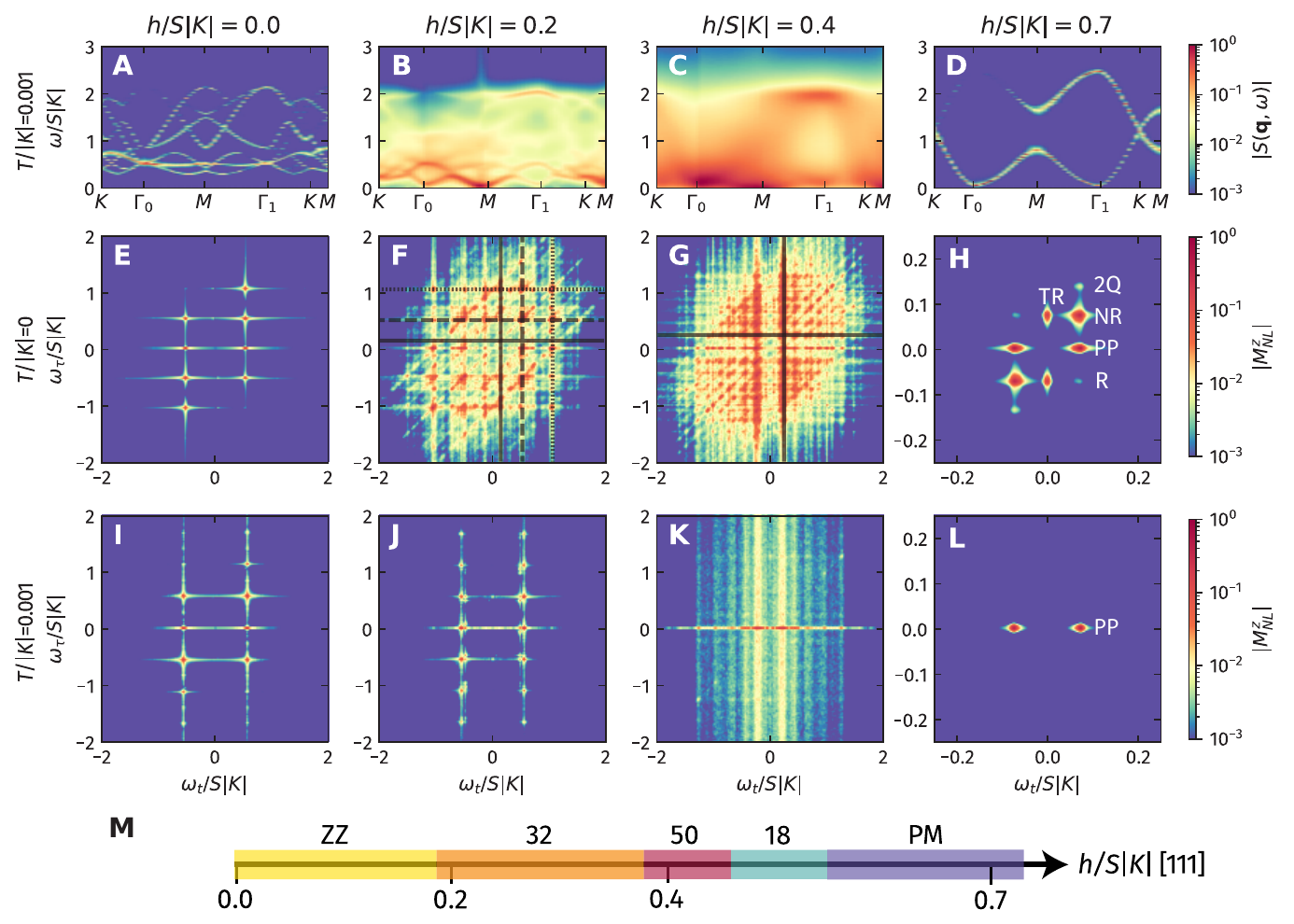}
    \caption{Field dependence of the dynamical spin structure factor (\textbf{A}-\textbf{D}) and two-dimensional coherent spectroscopy (\textbf{E}-\textbf{L}) for $K=-1$, $\Gamma=0.25$, and $\Gamma'=-0.02$. \textbf{A}-\textbf{D} are replotted from \citet{zhang_spin_2023}, and were performed at $T/|K|=0.001$. $|M_{NL}^{z}(\omega_t, \omega_\tau)|$  is shown for $T/|K|=0$ (\textbf{E}-\textbf{H}) and $T/|K|=0.001$ (\textbf{I}-\textbf{L}). Panel \textbf{F} includes black solid, dashed, and dotted lines at energies corresponding to the three high intensity peaks within the $\Gamma_0$ continuum in \textbf{B}. Panel \textbf{G} includes a solid black line at the energy corresponding to the high intensity peak in panel \textbf{C} at $\Gamma_0$. In panels \textbf{H} and \textbf{L}, TR=Terahertz Rectification, 2Q=2 Quantum, NR=Non-Rephasing, R=Rephasing, PP=Pump Probe. $\mathbf{M}$, The zero temperature classical phase diagram at each field. ZZ=Zig-Zag, 32=32-site, 50=50-site, 18=18-site, PM=Polarized paramagnet.}
    \label{fig:kggp}
\end{figure*}

Motivated by recent experiments using high magnetic fields to reveal a disordered quantum paramagnet at intermediate field strengths\cite{zhou_possible_2023}, we examine a minimal model of \aru at various magnetic fields. We choose the parameterization $(K,\Gamma,\Gamma')=(-1,0.25,-0.02)$, which yields large unit cell magnetic orders at intermediate fields between the zig-zag (ZZ) order at zero field and polarized phase at high field strengths. From the DSSF computed in previous studies for this parameter set plotted in Fig.~\ref{fig:kggp}A-D, a continuum of excitations can be seen at intermediate fields in panels B (32-site unit cell) and C (50-site unit cell). At finite temperatures, these fields host competing large unit cell orders which form a thermal ensemble, giving rise to continuum-like behavior\cite{franke_thermal_2022,zhang_spin_2023}. Fig.~\ref{fig:kggp}C in particular bears resemblance to the continuum seen in the pure ferromangetic Kitaev model in Fig.~\ref{fig:purekitaev}A. 
From the DSSF alone, it is hard to distinguish between this scenario and the continuum arising from the fractionalized excitations in the Kitaev spin liquid. 

We present the 2DCS for the $K\Gamma\Gamma'$ model at zero temperature in Figs.~\ref{fig:kggp}E-H and at finite temperature in Figs.~\ref{fig:kggp}I-L in the $z$ measurement channel. We observe sharp peaks at zero temperature appearing at energies corresponding to the coherent magnon branches at the $\Gamma_0$ point in the DSSF. Each peak corresponds to a distinct nonlinear optical process and will appear at frequencies with integer multiples of $E_n$, the energy of the $n^{\text{th}}$ magnon branch. These nonlinear processes are most clear in the polarized phase, where only the lower magnon branch contributes to the 2DCS response. For example, the peak at $(\omega_t,\omega_\tau)=(E_{\text{lower}},0)$ is the so-called ``pump-probe" response (labelled in Figs.~\ref{fig:kggp}H and L by ``PP"). In principle, each magnon band may exhibit such nonlinear signals, as described in detail in \citet{lu_coherent_2017}, where they measured the 2DCS spectra for the isolated magnon modes of the canted antiferromagnet YFO. Thus, for larger unit cell orders, we expect the 2DCS to be a collection of nonlinear signals for each magnon mode. 
For the ZZ phase, we see a collection of peaks in panel E at energies corresponding to the lower magnon branch from panel A. The intensity of the upper magnon branch is much weaker than the lower one in the DSSF, thus the 2DCS is dominated by the latter. For the 32-site and 50-site orders, each excitation on the continuum could in principle contribute peaks corresponding to all of the nonlinear processes described above, along with sum and difference frequency generation between the modes. As a result, the 2DCS exhibits a dense collection of many discrete peaks. Although a continuum is present, we can make out peaks in the 2DCS at energy scales corresponding to the strongest intensity peaks in the DSSF. These energies are shown in F as the black solid, dashed, and dotted lines for the three intense branches from B, while the highest intensity peak from C is marked by black solid lines in G. This dense collection of peaks can be directly contrasted with the continuous linear signals seen in the pure Kitaev model in Fig.~\ref{fig:purekitaev}. Thus, although both scenarios yield similar excitation continua in the DSSF, 2DCS produces qualitatively different results, reflecting the fundamentally distinct physics underlying the broad spectra. 

At finite temperature, in general we observe far fewer features than at zero temperature, similar to the pure Kitaev model case. The disappearance of the features present at zero temperature is particularly dramatic in panels J through L. For example, only the PP peak remains for the polarized phase, and the 2Q, TR, NR, and R peaks all disappear. However, these peaks are still visible in experiments done at finite temperature\cite{lu_coherent_2017,zhang_terahertz_2024}, whereas most of our peaks disappear at $T/|K|=0.001$. The fragility of the signals at finite temperature again demonstrates the sensitivity to thermal noise in our simulations, drawing attention to the role that quantum effects may play in stabilizing these signals in actual experiments. 

\section{Discussion}\label{sec:discussion}
To recap, we have demonstrated the strengths of 2DCS as a probe for the underlying excitations of exotic states emerging from frustrated spin systems. Using classical MD simulations on the pure Kitaev model, we observed sharp linear signals, notably diagonal nonrephasing and antidiagonal rephasing signals at zero temperature, reminiscent of the sharp linear features of the quantum model. Thus, the ability of purely classical dynamics to produce such signals implies that they alone cannot be taken as evidence of quantum spin liquid physics. Rather, additional details are required. Aside from the presence or absence of a gap, we presented additional discriminating features between the classical and quantum Kitaev models. First, we observed finite signals in measurement channels which should be forbidden in the quantum model due to the flux selection rules. Second, the classical model did not give rise to diagonal signals offset in $\omega_t$ and $\omega_\tau$, and was therefore not able to distinguish between the diagonal and shifted diagonal contributions in the quantum model. These two features are hallmarks of 2DCS on the exactly solvable Kitaev honeycomb model, and can be used to differentiate between the quantum Kitaev model and its classical counterpart. 

We also highlighted the usefulness of 2DCS in differentiating between distinct mechanisms that can give rise to spin excitation continua in INS. In the \kggp model, we previously reported a broad continuum at finite fields arising from the many excitations of an ensemble of large unit cell magnetic orders\cite{zhang_spin_2023}. Naively, this INS response looks similar to the one that arises from the classical Kitaev spin liquid state. However, using 2DCS, we observed a dense collection of discrete points in the nonlinear response of these large unit cell orders, in stark contrast with the sharp linear signals in the pure Kitaev model. Thus, while linear response fails to distinguish between excitation continua arising from spin liquid physics and large unit cell magnetic orders, 2DCS exhibits clear qualitative differences between the two cases.     

Our work establishes the ability of 2DCS to probe properties of excitations in spin systems otherwise inaccessible by linear spectroscopies, not only for classical versus quantum phenomena, but also for revealing the intrinsic nature of the excitations hidden in the broad spin excitation continua seen in INS. However, a few open questions remain. For example, what happens to the spectra in the quantum model in the presence of an external magnetic field and finite temperature, especially in the crossover region between a KSL and the polarized paramagnet? This question is relevant for distinguishing between a putative intermediate-field spin liquid and the field-polarized state in candidate Kitaev materials. Away from the exactly solvable point, the breaking of flux conservation negates the strict flux selection rules and hard flux gap,meaning that the quantum response should more closely resemble the classical case. This poses a challenge in identifying distinguishing characteristics of quantum spin liquid physics. Additionally, in both the pure Kitaev and \kggp models, the 2DCS simulated using classical MD are very sensitive to thermal noise, as many of the characteristic excitations of the underlying ground state disappear. The robustness of these signals in experiments \cite{lu_coherent_2017, zhang_terahertz_2024} may imply the importance of quantum effects in stabilizing them in the presence of thermal fluctuations. More generally, clarifying the role of quantum effects in 2DCS is thus another important step in understanding the nonlinear dynamics of frustrated magnets, and for quantum materials in general.

\begin{acknowledgments}
We thank Daniel J.~Schultz for helpful discussions. This work was supported by the Natural Science and Engineering Council of Canada (NSERC) Discovery Grant No. RGPIN-2023-03296 and the Center for Quantum Materials at the University of Toronto. Computations were performed on the Cedar cluster hosted by WestGrid and SciNet in partnership with the Digital Research Alliance of Canada.

\end{acknowledgments}

\appendix
\section{Details of the numerical methods}\label{sec:appmethods}

\begin{figure}[t!]
    \centering
    \includegraphics[width=1.0\columnwidth]{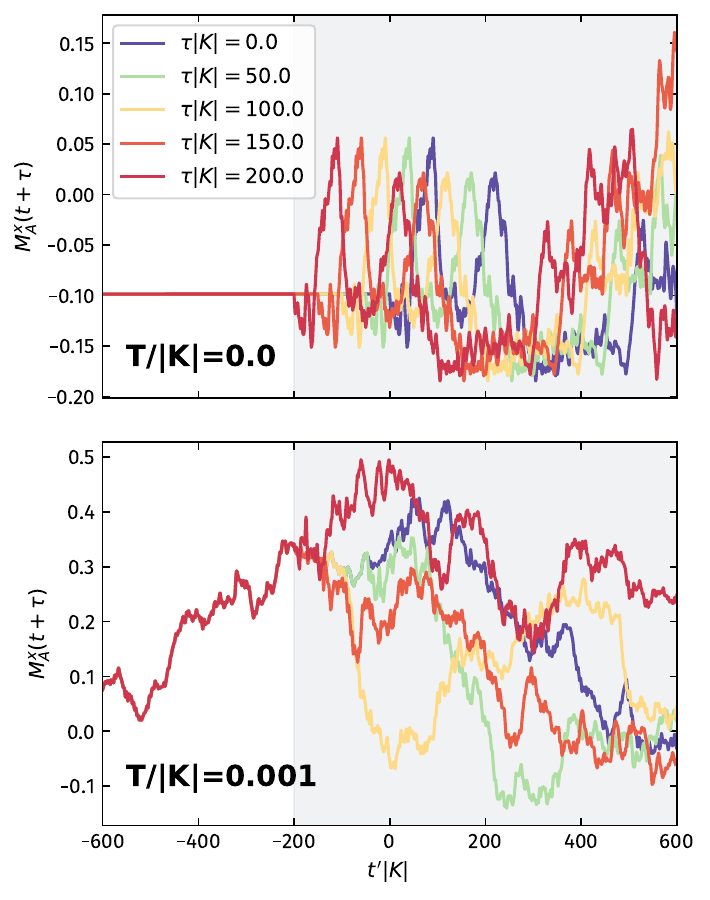}
    \caption{Magnetization at zero versus finite temperature in the presence of pulse A at various delay times $\tau$ for $K=-1$. The shaded region depicts the response after the pulse arrived. }
    \label{fig:appthermal}
\end{figure}
We used classical Monte Carlo (MC) techniques to obtain the spin configurations needed to compute the 2DCS. We implemented an adaptive Gaussian update method\cite{alzate-cardona_optimal_2019-1}, where each random spin sampling process will attempt to update within a width $\sigma$ away from the current spin, where $\sigma$ gets updated every sweep to converge towards an “optimal” $\sigma$. This optimal $\sigma$ is such that the target acceptance rate is approximately 50\%. We used this algorithm to study system sizes of up to $L=20\times20\times2$, and we thermalized 200 independent Markov chains to the desired temperature using $10^6$ MC sweeps via annealing. For faster convergence, we also used 100 overrelaxation sweeps for every MC sweep, where the spins are flipped along the local fields\cite{pixley_large-scale_2008}. For the zero temperature ($T/|K|=10^{-7}$) configurations, we performed $10^8$ deterministic update sweeps, where we randomly aligned the spins towards the local fields\cite{janssen_honeycomb-lattice_2016}. 

The spin configurations are then used as initial configurations (IC) for molecular dynamics, where each measurement is time-evolved deterministically according to the Landau-Lifshitz-Gilbert equations of motion as described in the main text\cite{lakshmanan_fascinating_2011-1}. For the input pulses, we used Gaussian pulse shapes described by $A(t)=\exp{[(-t/2t_0)^2]}\cos(2\pi f_0 t)$ where $t_0=0.38|K|^{-1}$ and $f_0=0.33|K|$. The strengths of the A and B pulses were $0.1|K|S$. For the numerical integration, we used the SSPRK53 ODE solver from the DifferentialEquations.jl Julia package\cite{rackauckas_differentialequationsjl_2017-1, zhang_maximum-principle-satisfying_2011}. We used a time window of $t_{\max}=600|K|^{-1}$ in timesteps of  $\delta t=0.25|K|^{-1}$ for the simulations. These results were then numerically Fourier transformed in the $t$ and $\tau$ dimensions to obtain the 2DCS spectra. 

\section{Role of thermal fluctuations in 2DCS from LLG}\label{sec:appthermal}

As mentioned in the main text, thermal noise can dramatically change the signals seen in 2DCS. This can be understood by looking at how the magnetization changes over time in the presence of a single pulse field, plotted in Fig.~\ref{fig:appthermal}. At zero temperature, the magnetization is completely flat before the pulse arrives, whereas at finite temperature, there are already non-trivial spin dynamics present. After the pulse arrives, at time $t^\prime=-\tau$, the signals at zero temperature for each $\tau$ are initially the same, just shifted by $\tau$, whereas the signals at finite temperature are dramatically different for each $\tau$. The presence of non-trivial dynamics, due to thermal fluctuations, already present in the absence of any pulses leads to the washing out of coherent signals in the resulting 2DCS spectra. 

\bibliography{ref}

\end{document}